\documentclass[prl,aps,twocolumn,preprintnumbers, showpacs, nofootinbib,superscriptaddress,notitlepage]{revtex4-1}
\usepackage{amssymb,amsthm,amsmath}
\usepackage{graphicx}   
\usepackage{color}      
\usepackage{slashed}    
\usepackage{epsfig}
\usepackage{subfigure}  
\usepackage{diagbox}
\usepackage[normalem]{ulem}

\begin{document}


\title{Lattice-QCD Calculations of TMD Soft Function Through \\ Large-Momentum Effective Theory}

\vspace{1.0cm}

\collaboration{\bf{Lattice Parton Collaboration ($\rm {\bf LPC}$)}}

\author{Qi-An Zhang}
\affiliation{Shanghai Key Laboratory for Particle Physics and Cosmology, MOE Key Laboratory for Particle Astrophysics and Cosmology, Tsung-Dao Lee Institute, Shanghai Jiao Tong University, Shanghai 200240, China}

\author{Jun Hua}
\affiliation{INPAC, Shanghai Key Laboratory for Particle Physics and Cosmology, MOE Key Laboratory for Particle Astrophysics and Cosmology, School of Physics and Astronomy, Shanghai Jiao Tong University, Shanghai 200240, China}

\author{Yikai Huo}
\affiliation{INPAC, Shanghai Key Laboratory for Particle Physics and Cosmology, MOE Key Laboratory for Particle Astrophysics and Cosmology, School of Physics and Astronomy, Shanghai Jiao Tong University, Shanghai 200240, China}
\affiliation{Zhiyuan College, Shanghai Jiao Tong University, Shanghai 200240, China}

\author{Xiangdong Ji}
\affiliation{Shanghai Key Laboratory for Particle Physics and Cosmology, MOE Key Laboratory for Particle Astrophysics and Cosmology, Tsung-Dao Lee Institute, Shanghai Jiao Tong University, Shanghai 200240, China}
\affiliation{Department of Physics, University of Maryland, College Park, MD 20742, USA}

\author{Yizhuang Liu}
\affiliation{Shanghai Key Laboratory for Particle Physics and Cosmology, MOE Key Laboratory for Particle Astrophysics and Cosmology, Tsung-Dao Lee Institute, Shanghai Jiao Tong University, Shanghai 200240, China}

\author{Yu-Sheng Liu}
\affiliation{Shanghai Key Laboratory for Particle Physics and Cosmology, MOE Key Laboratory for Particle Astrophysics and Cosmology, Tsung-Dao Lee Institute, Shanghai Jiao Tong University, Shanghai 200240, China}

\author{Maximilian Schlemmer}
\affiliation{Institut f\"ur Theoretische Physik, Universit\"at Regensburg, D-93040 Regensburg, Germany}

\author{Andreas Sch\"afer}
\affiliation{Institut f\"ur Theoretische Physik, Universit\"at Regensburg, D-93040 Regensburg, Germany}

\author{Peng Sun}
\affiliation{Nanjing Normal University, Nanjing, Jiangsu, 210023, China}

\author{Wei Wang}
\email{Corresponding author: wei.wang@sjtu.edu.cn}
\affiliation{INPAC, Shanghai Key Laboratory for Particle Physics and Cosmology, MOE Key Laboratory for Particle Astrophysics and Cosmology, School of Physics and Astronomy, Shanghai Jiao Tong University, Shanghai 200240, China}

\author{Yi-Bo Yang}
\email{Corresponding author: ybyang@itp.ac.cn}
\affiliation{CAS Key Laboratory of Theoretical Physics, Institute of Theoretical Physics, Chinese Academy of Sciences, Beijing 100190, China}
\affiliation{School of Fundamental Physics and Mathematical Sciences, Hangzhou Institute for Advanced Study, UCAS, Hangzhou 310024, China}
\affiliation{International Centre for Theoretical Physics Asia-Pacific, Beijing/Hangzhou, China}

\date{\today}

\begin{abstract}
The transverse-momentum-dependent (TMD) soft function is a key ingredient in QCD factorization of
Drell-Yan and other processes with relatively small transverse momentum.
We present a lattice QCD study of this function at moderately large rapidity
on a 2+1 flavor CLS dynamic ensemble with $a=0.098$ fm. We extract the rapidity-independent (or intrinsic) part of the 
soft function through a large-momentum-transfer pseudo-scalar
meson form factor and its quasi-TMD wave function using  leading-order factorization
in large-momentum effective theory.
We also investigate the rapidity-dependent part of the soft function---the Collins-Soper evolution kernel---based on the large-momentum evolution of the quasi-TMD wave function.
\end{abstract}
\maketitle

{\it Introduction.} For high-energy processes such as Higgs production at the Large-Hadron Collider,
quantum chromodynamics (QCD) factorization and parton distribution functions (PDFs) have been
essential for making theoretical predictions~\cite{Ellis:1991qj,Lin:2017snn}.
But for processes involving observation of a relatively small transverse momentum, $Q_\perp$
such as in Drell-Yan (DY) production and semi-inclusive deep inelastic scattering,
a new non-perturbative quantity called {\it soft function} is required  to capture
the physics of non-cancelling soft gluon-radiation at fixed $Q_\perp$~\cite{Collins:1981uk,Collins:1984kg,Ji:2004wu,Ji:2004xq}.
Physically, the soft function in DY is a cross section for a pair of a high-energy
quark and anti-quark (or gluon) traveling in the opposite light-cone directions
to radiate soft gluons of total transverse momentum $Q_\perp$ before they annihilate.
Although much progress has been made in calculating the soft function in perturbation
theory at $Q_\perp\gg \Lambda_{\rm QCD}$~\cite{Echevarria:2015byo,Li:2016ctv}, it is intrinsically non-perturbative when $Q_\perp$
is ${\cal O}(\Lambda_{\rm QCD})$.  Calculating the non-perturbative transverse-momentum-dependent (TMD)
soft function from first principles became feasible only recently~\cite{Ji:2019sxk}.

The main difference in such a calculation  in lattice QCD
is that it involves two light-like Wilson lines along directions
$n^\pm=\frac{1}{\sqrt 2}(1,\vec 0_\perp,\pm 1)$ in $(t,\perp,z)$ coordinates, making
direct simulations in Euclidean space impractical.  However,
much progress has been made in recent years in calculating physical quantities such as
light-cone PDFs using the framework of large-momentum effective theory (LaMET)~\cite{Ji:2013dva,Ji:2014gla}.
The key observation of LaMET is that the collinear  quark and gluon modes,
usually represented by light-like field correlators~\cite{Collins:2011zzd,Bauer:2000yr,Bauer:2001ct,Bauer:2001yt}, can be
accessed for large-momentum hadron states. A detailed review of LaMET
and its applications to collinear PDFs and other light-cone distributions
can be found in Refs.\cite{Ji:2020ect,Cichy:2018mum}. More recently,
some of the present authors have proposed that the TMD soft function
can be extracted from a special large-momentum-transfer form factor of
either a light meson or a pair of quark-antiquark color sources~\cite{Ji:2019sxk}.
Once calculated, the TMD factorization of the Drell-Yan and similar processes
can be made with entirely lattice-QCD-computable non-perturbative quantities~\cite{Ji:2014hxa,Ji:2018hvs,Ebert:2018gzl,Ebert:2019okf,Ji:2019ewn,Vladimirov:2020ofp}.

The TMD soft function is often defined and applied not in momentum space but in transverse coordinate space in terms of  the Fourier transformation variable $b_\perp $. In addition, it also depends on the ultraviolet (UV) renormalization scale $\mu$ (often defined
in dimensional regularization and minimal subtraction or $\overline{\rm MS}$) and rapidity regulators $Y+Y'$
\cite{Collins:2011zzd,Ji:2019sxk},
\begin{align}
S(b_\perp,\mu,Y+Y')=e^{(Y+Y')K(b_\perp,\mu)}S_I^{-1}(b_\perp, \mu)
\end{align}
where the first factor is related to rapidity evolution [described by the Collin-Soper (CS) kernel $K$], and the
second factor $S_I$ is the intrinsic, rapidity independent, part of the soft contribution.
The rapidity-regulator-independent CS-kernel $K$ is found calculable by taking ratio of
the quasi-TMDPDF at two different momenta~\cite{Ebert:2018gzl,Ebert:2019okf,Ji:2019ewn,Vladimirov:2020ofp,Ebert:2019tvc,Shanahan:2020zxr}.
On the other hand, calculating  the intrinsic soft function on the lattice
has never been attempted before.

In this paper we present the first lattice QCD calculation of the intrinsic soft function $S_I$
with several momenta on a 2+1 flavor CLS ensemble with $a=0.098$~fm~\cite{Bruno:2014jqa}, see Table I. In particular we perform  simulations of the large-momentum light-meson form factor and quasi-TMD wave functions (TMDWFs), whose ratio gives the intrinsic soft function~\cite{Ji:2019sxk}. The Wilson loop matrix element will be  used  to remove the linear divergence in the quasi-TMD wave function. The CS kernel, $K$, can also be calculated from the external momentum dependence
of the quasi-TMD wave function~\cite{Ji:2020ect}, and
we will calculate it as a by-product. Our result is consistent with that of quenched lattice calculations of TMDPDFs~\cite{Shanahan:2020zxr}.
 
\begin{figure}[!th]
\begin{center}
\includegraphics[width=0.45\textwidth]{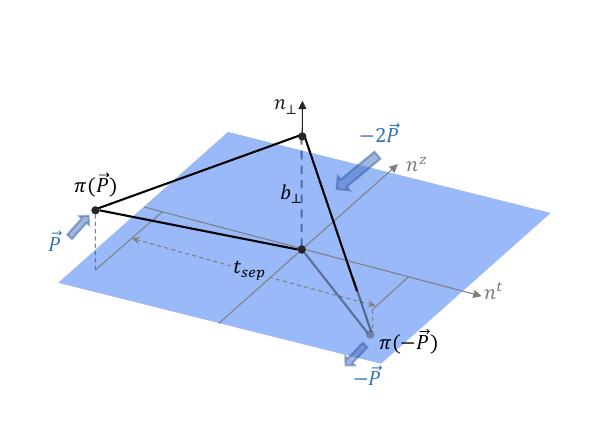}
\caption{Illustration of the pseudo-scalar meson  form factor $F$ calculated in this work.
The initial and final momenta of the pion are large and opposite. The transition
``current'' is made of two local operators at a fixed spatial separation $b_\perp$. $t_{\rm sep}$
is the time separation  between the source and sink of the pion.}\label{fig:formfactor}
\end{center}
\end{figure}

{\it Theoretical Framework.} The intrinsic soft function ($S_I$) can be obtained from the QCD
factorization of a large-momentum form factor of a non-singlet light pseudo-scalar meson with constituents $\pi=\overline{q}_2\gamma_5 q_1$, with the transition current made of two quark-bilinears 
with a fixed transverse separation $\vec{b}=(\vec{n}_\perp b_\perp, 0)$,
\begin{align}\label{eq:FF}
F(b_\perp,P^z)=\langle \pi(-\vec{P})|(\overline{q}_1\Gamma q_1)(\vec{b}) (\overline{q}_2\Gamma q_2)(0)|\pi(\vec P)\rangle_c. 
\end{align}
Here $q_{1,2}$ are light quark fields of different flavors, and $\vec{P}=(\vec{0}_{\perp},P^z)$.  {To extract the soft-factor, {operators and mesonic states are chosen such that each of the four lines in Fig.~\ref{fig:formfactor} are of a different
flavor as pointed out in Ref.~\cite{Ji:2019sxk}.}. The simplest scenario would correspond to the contraction in Fig.~\ref{fig:formfactor}, which shares the same topology as the so-called connected insertion. Thus  a subscript $c$ is added on the right-hand side of Eq.~(\ref{eq:FF}). By construction, the disconnected insertion   is not relevant  in this scenario which we will adopt in this work. }

It can be shown that the form factor defined in Eq.~(\ref{eq:FF}) is factorizable into the quasi-TMDWF $\Phi$ and the intrinsic soft function $S_I$~\cite{Ji:2019sxk,Ji:2020ect}
\begin{align}\label{eq:FF_factorization}
&F(b_\perp,P^z)= {S_I(b_\perp)}\\ &\times {\int_0^1 dx\, dx' H(x,x',P^z)\Phi^{\dagger}(x',b_\perp,-P^z)\, \Phi(x,b_\perp,P^z)}\nonumber
\end{align}
where $H$ is {the} perturbative hard kernel.
The quasi-TMDWF $ \Phi$ is the Fourier transformation of the coordinate-space correlation function
\begin{align}
&\phi(z,b_\perp,P^z)=\lim_{\ell\to\infty}\frac{\phi_\ell(z,b_\perp,P^z,\ell)}{\sqrt{Z_E(2\ell,b_\perp)}},\\
&\phi_\ell(z,b_\perp,P^z,\ell)\nonumber\\
&=\Big\langle 0\Big|\overline q_1\left(\frac{z }{2}n^z +\vec b\right)\Gamma_\Phi\,{\cal W}(\vec b,\ell)q_2\left(-\frac{z}{2}n^z \right)\Big| {\pi(\vec{P})}\Big\rangle\nonumber. 
\end{align}
In the above ${\cal W}(\vec b,\ell)$ is the spacelike staple-shaped gauge link,
\begin{align}
{\cal W}(\vec b,\ell)&= {\cal P}{\rm exp} \left[ ig_s\int_{-\ell}^{z/2} \textrm{d}s\ n^z\cdot  A(n^z s+b_\perp)\right] \nonumber\\
& \times {\cal P}{\rm exp} \left[ ig_s\int_{0}^{b_\perp } \textrm{d}s\ n_\perp\cdot  A(-\ell n^z +s  n_\perp)\right] \nonumber\\
& \times {\cal P}{\rm exp} \left[ ig_s\int_{-z/2}^{-\ell} \textrm{d}s\ n^z\cdot  A(n^z s)\right],
\end{align}
$n^z$ and $n_\perp$ are the unit vectors in $z$ and  transverse directions respectively. 
$Z_E(2\ell,b_\perp)$ is the vacuum expectation value of a rectangular spacelike Wilson loop with size $2\ell \times b_\perp$ which   removes the pinch-pole singularity and Wilson-line self-energy in quasi-TMDWF~\cite{Ji:2019sxk}.

Since the UV divergence of the intrinsic soft function is multiplicative~\cite{Ji:2020ect},
the  ratio  $S_I(b_{\perp},1/a)/S_I(b_{\perp,0},1/a)$ calculable on lattice is UV
renormalization-scheme independent, where $b_{\perp,0}$ is a reference distance which is taken
small enough to be calculated perturbatively. Thus we
can obtain the result in the $\overline{\rm MS}$ scheme
through
\begin{align}\label{eq:S_ratio}
S_{I,\overline{\rm MS}}(b_\perp,\mu)= \left(\frac{S_I(b_{\perp},1/a)}{S_I(b_{\perp,0},1/a)}\right)S_{I,\overline{\rm MS}}(b_{\perp,0},\mu)
\end{align}
where $S_{I,\overline{\rm MS}}(b_{\perp,0},\mu)$ is perturbatively calculable, e.g.,
\begin{align}\label{eq:S_ratio_1loop}
S_{I,\overline{\rm MS}}(b_{{\perp}},\mu)=1-\frac{\alpha_sC_F}{\pi}\ln\frac{\mu^2 b_{{\perp}}^2}{4 e^{-2\gamma_E}}+{\cal O}(\alpha_s).
\end{align}

In the present exploratory study, we will consider only   leading
order matching in Eq. (\ref{eq:FF_factorization}), for which the perturbative
kernel is $H(x,x',P^z)=1/({2N_c})+{\cal O}(\alpha_s)$, independent of $x$ and $x'$.
Using $\phi(0,b_\perp,-P^z)=\phi(0,b_\perp,P^z)$ under parity transformation, we obtain
\begin{align}\label{eq:S_I_z=0}
S_I(b_\perp)=\frac{2N_c F(b_\perp,P^z)}{|\phi(0,b_\perp,P^z)|^2}+{\cal O}(\alpha_s, (1/P^z)^2),
\end{align}
where power corrections from finite $P^z$ are ignored. Since $P^z$ is related
to the rapidity of the meson, we henceforth replace it by the boost factor $\gamma\equiv E_{\pi}/m_{\pi}$.
Eq.~(\ref{eq:S_ratio}) can be written as
\begin{align}\label{eq:S_ratio_z=0}
S_{I,\overline{\rm MS}}(b_\perp,\mu)&=\frac{F(b_\perp,P^z)}{F(b_{\perp,0},P^z)}\frac{|\phi(0,b_{\perp,0},P^z)|^2}{|\phi(0,b_\perp,P^z)|^2}\nonumber\\
&+{\cal O}(\alpha_s, \gamma^{-2})\, .
\end{align}
The ratio on the right-hand side of the above expression  is independent of the renormalization scale $\mu$ since only the leading-order contribution is kept.

{On the other hand, the quasi-TMDWF} can be used to extract the Collins-Soper kernel $K$ using a method similar
to~\cite{Ebert:2018gzl}
\begin{align}
&K(b_\perp,\mu)=\frac{1}{\ln(P_1^z/P_2^z)}\ln\left|\frac{C(xP_2^z,\mu) \Phi_{\overline{\rm MS}}(x,b_\perp,P_1^z,\mu)}{C(xP_1^z,\mu) \Phi_{\overline{\rm MS}}(x,b_\perp,P_2^z,\mu)}\right|\label{eq:CS_kernel}\\
&\quad=\frac{1}{\ln(P_1^z/P_2^z)}\ln\left|\frac{\int_0^1 \textrm{d}x \Phi(x,b_\perp,P_1^z)}{\int_0^1 \textrm{d}x \Phi(x,b_\perp,P_2^z)}\right|+{\cal O}(\alpha_s, \gamma^{-2})\nonumber\\
&\quad=\frac{1}{\ln(P_1^z/P_2^z)}\ln\left|\frac{\phi(0,b_\perp,P_1^z)}{\phi(0,b_\perp,P_2^z)}\right|+{\cal O}(\alpha_s, \gamma^{-2})\label{eq:CS_kernel_z=0}. 
\end{align}
{In the second line,  again only the leading order matching kernel $C(xP^z,\mu)=1+{\cal O}(\alpha_s)$ is used. }
The  renormalization factors for $\Phi$ are cancelled.
The rapidity-scheme-independent CS kernel $K$ is independent of $\mu$ in this approximation because only the leading term has been kept.

While Eqs.~(\ref{eq:S_ratio}) and (\ref{eq:CS_kernel}) are exact and can be
used for precision {studies} in the future,  Eqs.~(\ref{eq:S_ratio_z=0}) and (\ref{eq:CS_kernel_z=0}) are the leading-order approximation used in this pioneering  work.

\begin{table}[htbp]
\begin{center}
\caption{\label{table:cls} Parameters used  in the numerical simulation. The first row shows the parameters of the 2+1 flavor clover fermion CLS ensemble (named A654) and the second one shows the number of the A654 configurations and valence pion mass used for this calculation.}
\begin{tabular}{ccccccc}
\hline
 $\beta$ & $L^3\times T$  &a (fm)  & $c_{sw}$  & $\kappa^{\rm sea}_l$ & $m^{\rm sea}_{\pi} $(MeV)\\
  3.34 & $24^3 \times  48$ & 0.098 &  2.06686  & 0.13675 & 333 \\
\hline
&&& $N_{cfg}$  &  $\kappa^{v}_l$  & $m^v_{\pi} $ (MeV)  \\
  &&&  864  & 0.13622   & 547 \\
 \hline
\end{tabular}
\end{center}
\end{table}

{\it Simulation setup.} For the present study, we use configurations generated with   2+1 flavor clover fermions and tree-level Symanzik gauge action configuration by the CLS collaboration using periodic boundary conditions~\cite{Bruno:2014jqa}. The detailed parameters are listed in Table~\ref{table:cls}. Note that $m_\pi=547$ MeV instead of 333 MeV is used for valence quarks in order to have a better signal. Physically,  the soft function
becomes  independent of the meson mass for large boost factors $\gamma$.

To calculate the form factor in Eq.(2), we generate the wall source propagator,
\begin{align}
 S_w(x,t,t';\vec{p})=\sum_{\vec{y}} S(t,\vec{x};t',\vec{y})e^{i\vec{p}\cdot(\vec{y}-\vec{x})},
\end{align}
on the Coulomb gauge fixed configurations at $t'=0$ and $t_{\rm sep}$ for both the initial and final meson states.  $S$ is the quark propagator from $(t',\vec{y})$ to $(t,\vec{x})$. Then we can construct the three point function (3pt)
corresponding to the form factor in Eq. (2),
\begin{align}
&C_3(b_\perp,P^z;p^z,t_{\rm sep},t)\\
&=\frac{1}{L^3} \sum_x\textrm{Tr} \langle  S_{w}^\dagger(\vec{x}+\vec{b},t,0;-\vec{p})\gamma_5\Gamma S_{w}(\vec{x}+\vec{b},t,t_{\rm sep};\vec{p}) \nonumber\\
&\;\;\quad\quad \quad \times S_{w}^\dagger(\vec{x},t,t_{\rm sep};-\vec{P}+\vec{p}) \gamma_{5}\Gamma
 S_{w}(\vec{x},t,0;\vec{P}-\vec{p})\rangle\nonumber.
\end{align}
The quark momentum $\vec{p}=(\vec 0_{\perp},p^z)$, and
the relation $\gamma_5 S^{\dagger}(x,y)\gamma_5=S(y,x)$ have been applied for the anti-quark propagator. We have tested several choices of $\Gamma$, and will use the unity Dirac matrix  $\Gamma=I$ as it has the best signal and describes the leading twist light-cone contribution in the large $P^z$ limit. Notice that the $\Gamma=\gamma_4$ case  is   subleading in the large $P^z$ limit and is less suitable,   although  the excited state contamination might be smaller.

By generating the wall source propagators at all the 48 time slices with quark momentum ${p}^z=(-2,-1,0,1,2) \times 2\pi/(La)$, we can maximize the statistics of the 3pt function with all the meson momenta $P_z$ from 0 to $8\pi/(La)$ ($\sim 2.1$ GeV) with arbitrary $t$ and $t_{\rm sep}$.  $C_3(b_\perp,P^z,t_{\rm sep},t)$ is related to the bare $F(b_\perp,P^z)$ using standard parameterization of 3pt with one excited {state},
\begin{align}
& C_3(b_\perp,P^z;p^z,t_{\rm sep},t)=\frac{A_w(p_z)^2}{(2E)^2}e^{-Et_{\rm sep}}\big[F(b_\perp,P^z)\nonumber\\
&\quad\quad+c_1(e^{-\Delta E t}+e^{-\Delta E (t_{\rm sep}-t)})+c_2e^{-\Delta E t_{\rm sep}}\big].
 \label{eq:3pt}
\end{align}
$A_w$ is the matrix element of the Coulomb gauge fixed wall (CFW) source pion interpolation field, $E=\sqrt{m_{\pi}^2+{P}^{z2}}$ is the pion energy, $\Delta E$ is the mass gap between pion and its first excited state, $c_{1,2} $ are parameters for the excited state contamination. Note that the $p_z$ dependence factor $A_w^2$ will cancel{.}

The same wall source propagators can be used to calculate the two-point function related to the bare quasi-TMDWF,
\begin{align}
&C_2(b_\perp,P^z;p_z,\ell,t)=\frac{1}{L^3\sqrt{Z_E(2\ell,b_\perp)}}\sum_x \textrm{Tr}e^{i\vec{P}\cdot \vec{x}}\nonumber\\
&\times\langle  S_w^\dagger(\vec{x}+\vec{b},t,0;-\vec{p}){\cal W}(\vec b,\ell) \gamma_5\Gamma_\Phi  S_w(\vec{x},t,0;P^z-\vec{p}) \rangle\nonumber\\
& =\frac{A_w(p_z)A_p}{2E}e^{-Et}\phi_\ell(0,b_\perp,P^z,\ell)(1+c_0 e^{-\Delta E t}),\label{eq:2pt}
\end{align}
where again we parameterize the mixing with one excited state. 
$A_p$ is the matrix element of the point sink pion interpolation field. It will be  removed when we normalize $\phi_\ell(0,b_\perp,P^z,\ell)$ with $\phi_\ell(0,0,P^z,0)$. We choose $\Gamma_\Phi=\gamma^t\gamma_5$
to define the wave function amplitude in Eq. (4). Based on the quasi-TMDPDF study in Ref.{~\cite{Shanahan:2019zcq,Shanahan:2020zxr}} with a similar staple-shaped gauge link operator, the mixing effect {could be sizable  when summing various contributions. In the supplemental material, we report a similar simulation but using the A654 ensemble. We find that the mixing effects can reach   order $5\%$ for the transverse separation $b_\perp\sim 0.6{\rm fm}$. These effects will be included in the following analysis as  one of the systematic uncertainties, while a comprehensive study on the mixing effects will be conducted in the future. }

\begin{figure}[!th]
\includegraphics[width=0.45\textwidth]{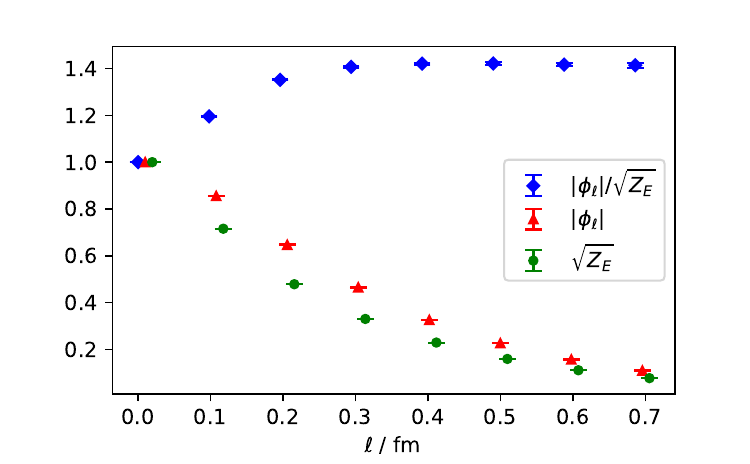}
\caption{ Results for the $\ell$ dependence of the  quasi-TMDWF with $z=0$, and also the square root of the Wilson loop which is  used for the subtraction, taking the $\{P^z, b_{\perp}, t\}=\{6\pi/L, 3a, 6a\}$ case as a example. All the results are normalized with their values at $\ell=0$.}\label{fig:L_dependence}
\end{figure}

The dispersion relation of the pion state, statistical checks for the  measurement histogram, and information on the autocorrelation between configurations can be found in the supplemental materials~\cite{supplemental}.

{\it Numerical Results.} Fig.~\ref{fig:L_dependence} shows  the   dependence of the norm of quasi TMDWFs on the length $\ell$ of the Wilson-line.  As one can see from this figure, with $\{P^z, b_{\perp}, t\}=\{6\pi/L, 3a, 6a\}$, both the quasi-TMDWF $\phi_\ell(0,b_\perp,P^z,\ell)$   and  the square root of the Wilson loop $Z_E$ decay exponentially with   length $\ell$, but the subtracted quasi-TMDWF is length independent when $\ell\ge 0.4$ fm. Some other cases with larger $P^z$, $b_{\perp}$, and $t$ can be found in the supplemental materials~\cite{supplemental}. Based on this observation, we will use $\ell=7a=0.686$ fm as asymptotic results for all cases in the following calculation.

\begin{figure}[!th]
\begin{center}
\includegraphics[width=0.45\textwidth]{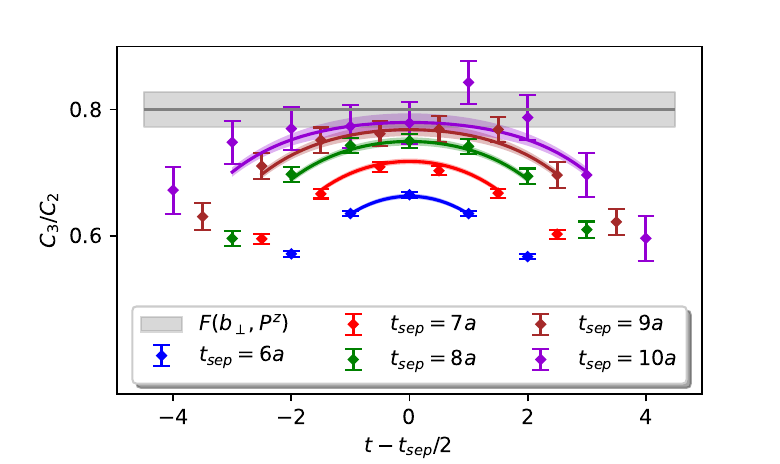}
\caption{The ratios $C_3(b_\perp,P^z,t_{\rm sep},t)/C_2(0, P^z,0,t_{\rm sep})$ (data points) which converge to the ground state contribution at $t,t_{\rm sep}\rightarrow \infty$ (gray band) as   function of $t_{\rm sep}$ and $t$, with $\{P^z, b_{\perp}\}=\{6\pi/L, 3a\}$. As in this figure, our data in general agree with the predicted fit function (colored bands). }\label{fig:joint_fit}
\end{center}
\end{figure}

We performed a joint fit of the form factor and quasi-TMDWF with the same $P^z$ and  $b_{\perp}$ with the parameterization in Eqs.~\eqref{eq:3pt} and \eqref{eq:2pt}. The ratios $C_3(b_\perp,P^z,t_{\rm sep},t)/C_2(0, P^z,0,t_{\rm sep})$ with different $t_{\rm sep}$ and $t$ for the $\{P^z, b_{\perp}\}=\{6\pi/L, 3a\}$ case are shown in Fig.~\ref{fig:joint_fit}, with ground state contribution (gray band) and the fitted results at finite $t_{2}$ and $t$ (colored bands). In this calculation, the excited state contribution is properly described by the fit with  $\chi^2/{\rm d.o.f.}=0.6$. The details of the joint fit, and also more fit quality checks are shown in the supplemental materials~\cite{supplemental}, with similar fitting quality.

\begin{figure}[!th]
\includegraphics[width=0.45\textwidth]{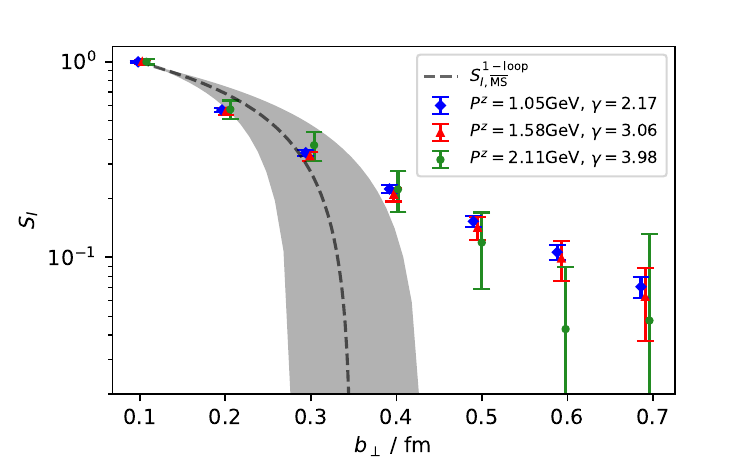}
\caption{ The intrinsic soft factor as a function of $b_{\perp}$  with  $b_{\perp,0}=a$ as in Eq.~\eqref{eq:S_ratio_z=0}. With different pion momentum $P^z$, the results are consistent with each other. The dashed  curve shows  the result of the 1-loop calculation, see Eq.~(\ref{eq:S_ratio_1loop}),  with the strong coupling constant $\alpha_s(1/b_{\perp})$. {The shaded band corresponds to the scale uncertainty  of $\alpha_s$: $\mu\in [1/\sqrt{2},\sqrt{2}]\times 1/b_\perp$. The systematic uncertainty from the operator mixing has been taken into account.  }}\label{fig:quasi-soft-factor}
\end{figure}

The resulting soft factor as function of $b_{\perp}$ is plotted in Fig.~\ref{fig:quasi-soft-factor}, at $\gamma$= 2.17,  3.06 and 3.98, which corresponds to $P^z=\{4,6,8\}\pi/L=\{1.05, 1.58, 2.11\}$ GeV respectively.
As in Fig.~\ref{fig:quasi-soft-factor}, the results at different large $\gamma$ are consistent with each other,
demonstrating that the asymptotic limit is stable within errors.
We also compare the intrinsic soft function extracted from the lattice to the one-loop result in Eq.~(\ref{eq:S_ratio_1loop}),
with $\alpha_s(\mu=1/b_\perp)$ evolving from $\alpha_s(\mu=2\;{\rm GeV})\approx 0.3$. {The shaded band corresponds to the scale uncertainty  of $\alpha_s$: $\mu\in [1/\sqrt{2},\sqrt{2}]\times 1/b_\perp$. }
Notice that the $b_\perp$ dependence of the former comes purely from the  lattice simulation, while that for the latter is from perturbation theory. 
{For ease of comparison, we also tabulate the results for the soft function in the supplemental material~\cite{supplemental}. }

\begin{figure}[!th]
\begin{center}
\includegraphics[width=0.45\textwidth]{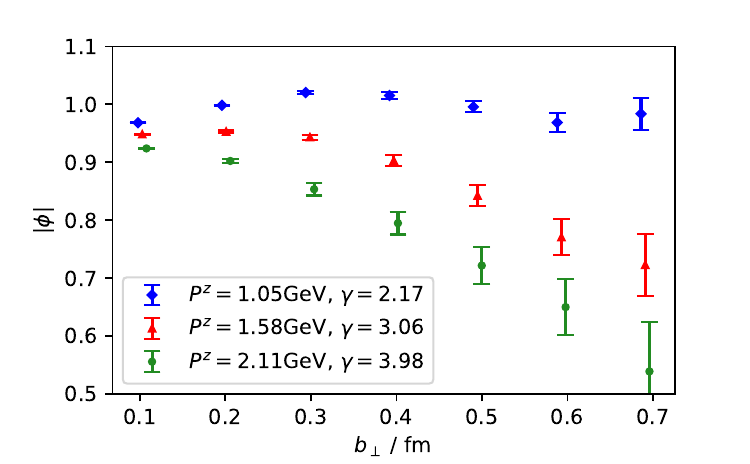}
\includegraphics[width=0.45\textwidth]{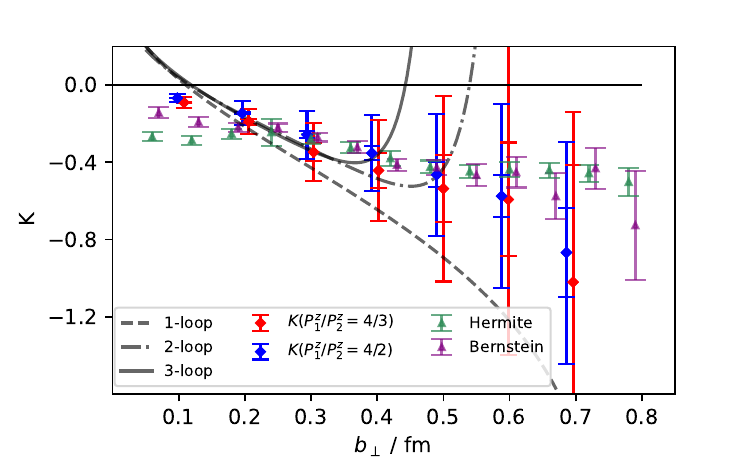}
\caption{ Quasi-TMDWF (upper panel) and extracted Collins-Soper kernel (lower panel), as functions of $b_{\perp}$. The visible  $P^z$ dependence of the quasi-TMDWF can be primarily  understood by that from the Collins-Soper kernel, as the kernel we obtained with tree level matching is consistent with up to 3-loop perturbative calculations (at small $b_{\perp}$) with the strong coupling $\alpha_s$ at the scale $1/b_{\perp}$, and also the non-perturbative result from the  pion quasi-TMDPDF. { Results based on quenched lattice calculations, labeled as ``Hermite" and ``Bernstein"~\cite{Shanahan:2020zxr},  are also shown for comparison. Errors in the lower panel correspond to the statistical errors and the systematic errors from the non-zero imaginary part as well as  the operator mixing effects.  }  }\label{fig:quasi-TMDWF}
\end{center}
\end{figure}

We can see a clear $P^z$ dependence in the quasi-TMDWF $|\phi_\ell(0,b_{\perp},P^z,\ell)|$ normalized with $\phi_\ell(0,0,P^z,0)$, as in the upper panel of Fig.~\ref{fig:quasi-TMDWF}. This dependence is related to the CS kernel as shown in Eq.~(\ref{eq:CS_kernel_z=0}), up to   possible LaMET matching effects and power corrections of order $1/\gamma^2$. Thus we use Eq.~(\ref{eq:CS_kernel_z=0}) to extract  the kernel in   the tree level approximation, and compare the result in the lower panel of Fig.~\ref{fig:quasi-TMDWF} with that of  Ref.~\cite{Shanahan:2020zxr} and up to 3-loop perturbative ones with $\alpha_s(\mu=1/b_\perp)$. { We estimate the systematic uncertainty by combining in quadrature the  contributions from the operator mixing effects, and   from the non-vanishing  imaginary part of the quasi-TMDWF which should be {cancelled by proper treatments on higher order effects}.} For details see the supplemental materials~\cite{supplemental}, in particular {Sec. C and F}. Our result is consistent with  that of Ref.~\cite{Shanahan:2020zxr}.

{\it Summary and Outlook.} In this work, we have presented an exploratory lattice calculation of the intrinsic
soft function by simulating the light-meson form factor of four-quark non-local operators and quasi-TMD wave functions. Our result shows a mild hadron momentum dependence, which allows a future precision study to eliminate the large momentum dependence using perturbative matching~{\cite{Ji:2020ect}}.
As a reliability check, the agreement between the CS kernel obtained from our quasi-TMDWF result  and   previous calculations shows that the systematic uncertainties {including the partially quenching effect,} the only leading  perturbative matching and missing power corrections $1/\gamma$ in LaMET expansion {might be} sub-leading. Our calculation paves the way towards the first principle predictions of   physical cross sections  for, e.g., Drell-Yan and Higgs productions at small transverse momentum.


\vspace{0.5cm}

{\it Acknowledgment.}---
We thank Xu Feng, Yuan Li, Shi-Cheng Xia, Jianhui Zhang and Yong Zhao for valuable discussions.
We thank the CLS Collaboration for sharing the lattice ensembles used to perform this study.
The LQCD calculations were performed using the Chroma software suite~\cite{Edwards:2004sx}.
The numerical calculation is supported by Chinese Academy of Science CAS Strategic Priority Research Program of Chinese Academy of Sciences, Grant No. XDC01040100,   HPC Cluster of ITP-CAS, and Jiangsu Key Lab for NSLSCS.
The  setup for numerical simulations was conducted  on the $\pi$ 2.0 cluster supported by the Center for High Performance Computing at Shanghai Jiao Tong University. 
J.~Hua is supported by NSFC under grant No. 11735010 and 11947215.
Y.-S. Liu is supported by National Natural Science Foundation of China under grant No.11905126.
M.~Schlemmer  and A.~Sch\"afer were supported by the cooperative research center CRC/TRR-55 of DFG. 
P. Sun is supported by Natural Science Foundation of China under grant No. 11975127 as well as Jiangsu Specially Appointed Professor Program.
W. Wang is supported in part by Natural Science Foundation of China under grant No.   11735010, 11911530088,  by Natural Science Foundation of Shanghai under grant No. 15DZ2272100.
Q.-A. Zhang is supported by the China Postdoctoral Science Foundation and the National Postdoctoral Program for Innovative Talents (Grant No. BX20190207).

\clearpage

\begin{widetext}
\section*{Supplemental Materials}

\subsection{Simulation checks}

\begin{figure}[!th]
\begin{center}
\includegraphics[width=0.5\textwidth]{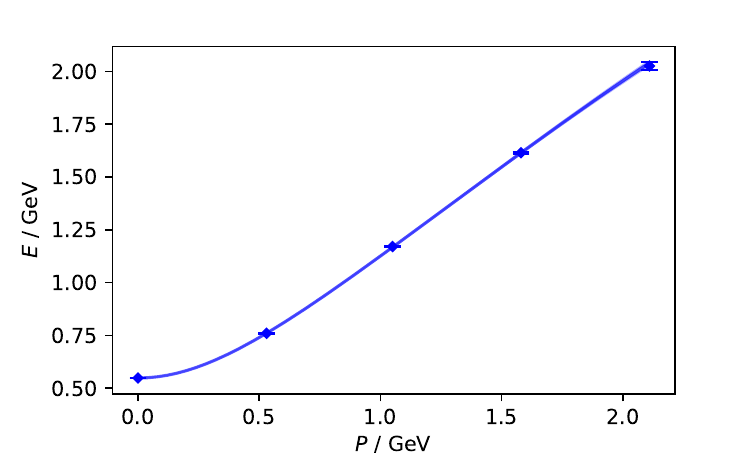}
\caption{ The dispersion relation of the pion state with the pion mass  from the 2pt function. The data up to 8$\pi/L$ ($\sim$2 GeV) can be described with the formula $E_{\pi}=\sqrt{m_{\pi}^2+c_1 P^2+c_2P^4a^2}$ with $c_1=0.9945(40)$ and $c_2=-0.0282(27)$. The deviation at 8$\pi/L$ from the continuum limit is around 2\%.
}\label{fig:dispersion}
\end{center}
\end{figure}

Fig.~\ref{fig:dispersion} shows the dispersion relation with the pion mass we used. The curve  shows the fit based on the formula $E_{\pi}=\sqrt{m_{\pi}^2+c_1 P^2+c_2P^4a^2}$, where the last term in the square root parameterizes discretization errors.   We used   momenta up to  8$\pi/L$ ($\sim$2 GeV).  The fit gives results---$c_1=0.9945(40)$ and $c_2=-0.0282(27)$--- that are consistent with the ground state energy  calculated from two point function.  It indicates only small discretization errors.Thus it is  expected that the dispersion relation can  recover the standard $E_{\pi}=\sqrt{m_{\pi}^2+P^2}$ in the continuum limit.

\begin{figure}[!th]
\begin{center}
\includegraphics[width=0.4\textwidth]{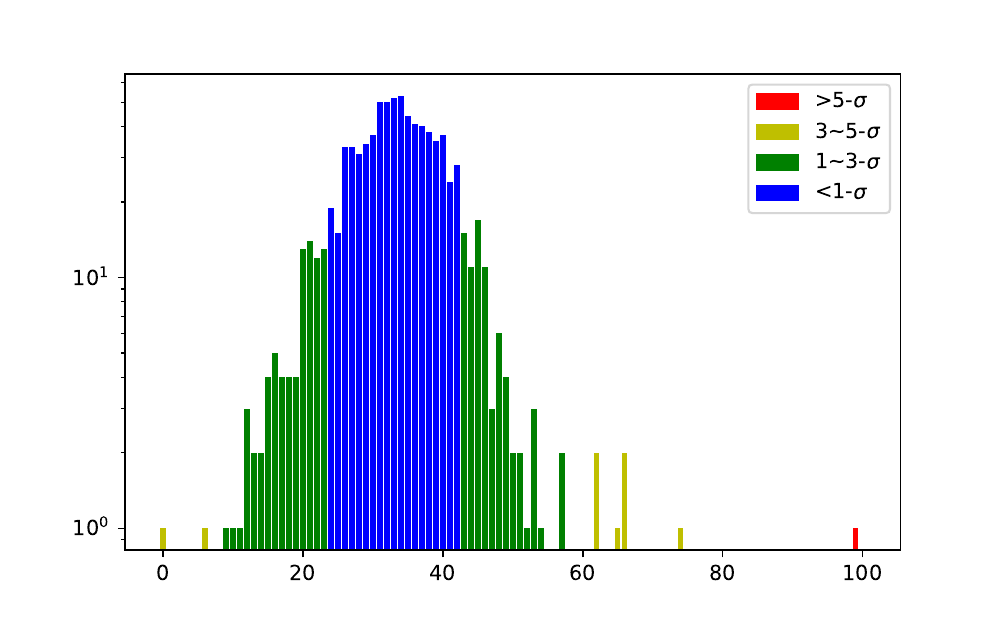}
\includegraphics[width=0.4\textwidth]{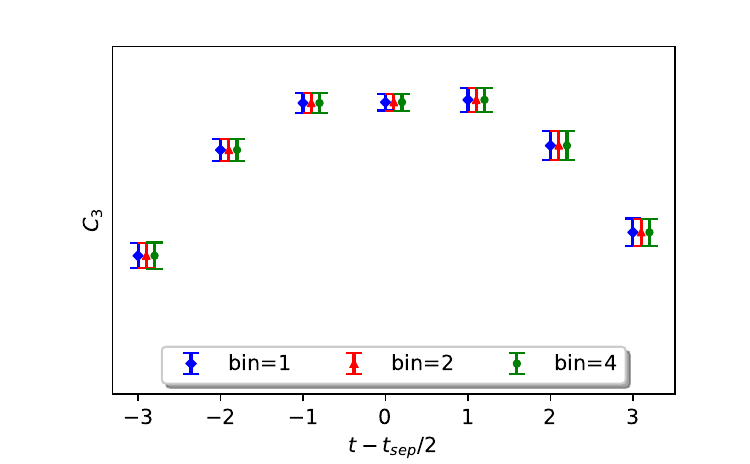}
\caption{Statistical check on the simulation, taking the form factor  with $P^z=6\pi/L$, $b_{\perp}$=3, $t_{2}$=8 and $t=t_{2}/2$ as example. The left panel shows the histogram for {$864$ configurations which are included
in the further analysis}, and the right panel shows the bin size dependence after we averaged all the measurements on each configuration.}\label{fig:statistics}
\end{center}
\end{figure}

Taking the form factor  with $P^z=6\pi/L$, $b_{\perp}$=3, $t_{2}$=8 and $t=t_{2}/2$ as example, Fig.~\ref{fig:statistics} shows the statistical check of the measurements we did. {We analysed 868 configurations and dropped 4 of them in the analysis due to very strong localized artifacts.} The left panel shows the histogram of 864 (configurations) $\times$ 48 (time {slices}) measurements. {It has been noticed that using the clover action with light mass and/or a coarse lattice on the dynamical configuration, the exceptional measurement, though very rare, can occur since the critical point  is not very stable. It turns out that  some  strongly localized artifacts were not observed in other CLS ensembles with finer lattice spacings but can happen in a few configurations of the coarse ensembles, for example the  A654 ensemble which we used in this analysis.  Since this is  a small portion of the total configurations, namely $4/868$, removing these configurations might be plausible. }

 After we average the measurements over the same configuration, we find that the autocorrelation effect is negligible, since no obvious bin size dependence of the result is observed, as shown in the right panel of Fig.~\ref{fig:statistics}.

\subsection{$\ell$ dependence of TMDWF}

\begin{figure}[!th]
\begin{center}
\includegraphics[width=0.9\textwidth]{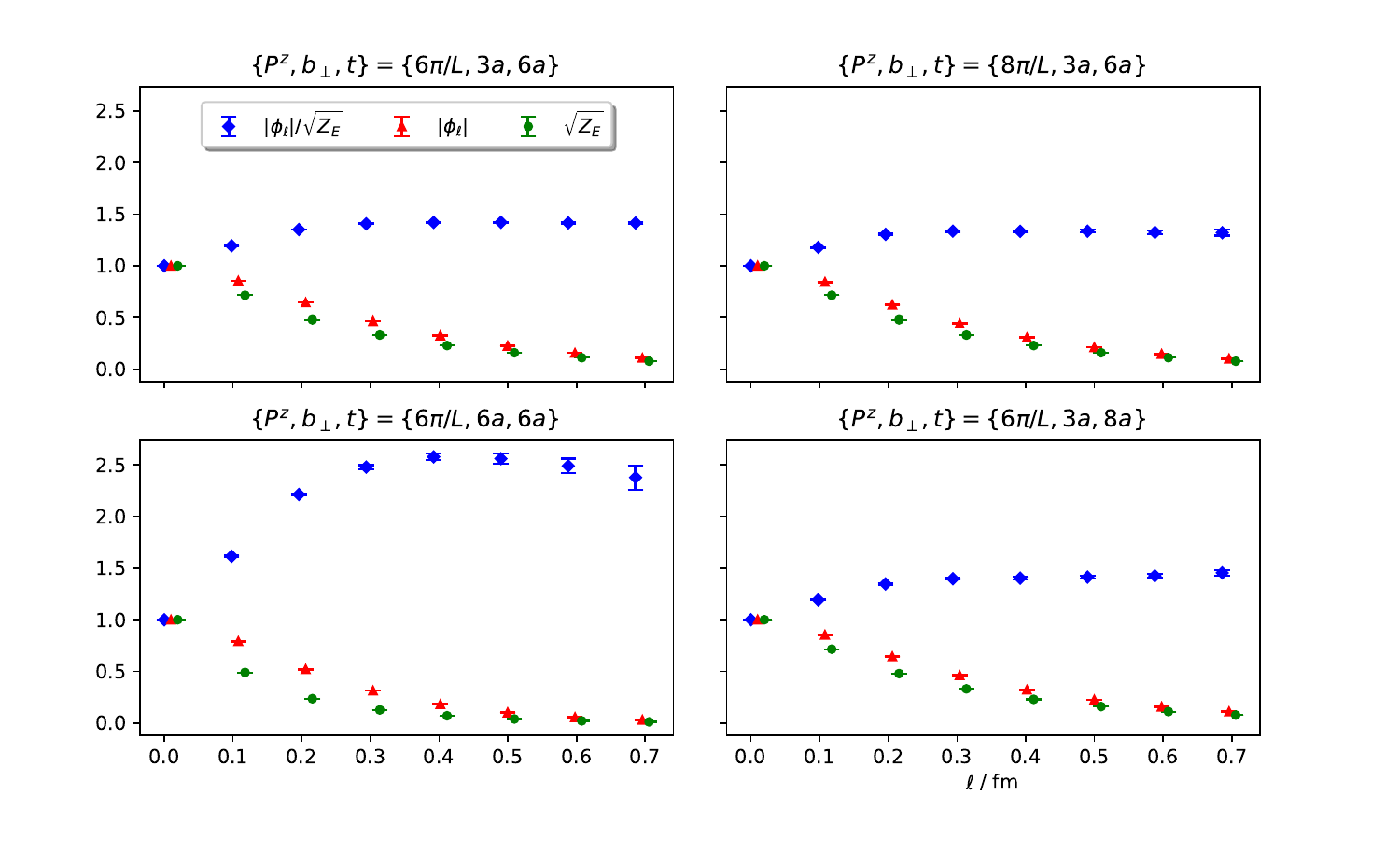}
\caption{Figure for the $\ell$ dependence of $|\phi_\ell(0,b_\perp,P^z,\ell)|$ with $\{P^z, b_{\perp}, t\}=\{6\pi/L, 3a, 6a\}$ (top left, the case shown in Fig.~2), $\{P^z, b_{\perp}, t\}=\{8\pi/L, 3a, 6a\}$ (top right), $\{P^z, b_{\perp}, t\}=\{6\pi/L, 6a, 6a\}$ (bottom left), and $\{P^z, b_{\perp}, t\}=\{6\pi/L, 3a, 8a\}$ (bottom right).}\label{fig:L_dependence2}
\end{center}
\end{figure}


In Fig.~\ref{fig:L_dependence2}, we give the $\ell$ dependence of $|\phi_\ell(0,b_\perp,P^z,\ell)|$ for a few more cases, similar to the $\{P^z, b_{\perp}, t\}=\{6\pi/L, 3a, 6a\}$ case shown in Fig.~2 but with larger $P^z$, $b_{\perp}$ and also $t$.

\begin{figure}[!th]
\begin{center}
\includegraphics[width=0.45\textwidth]{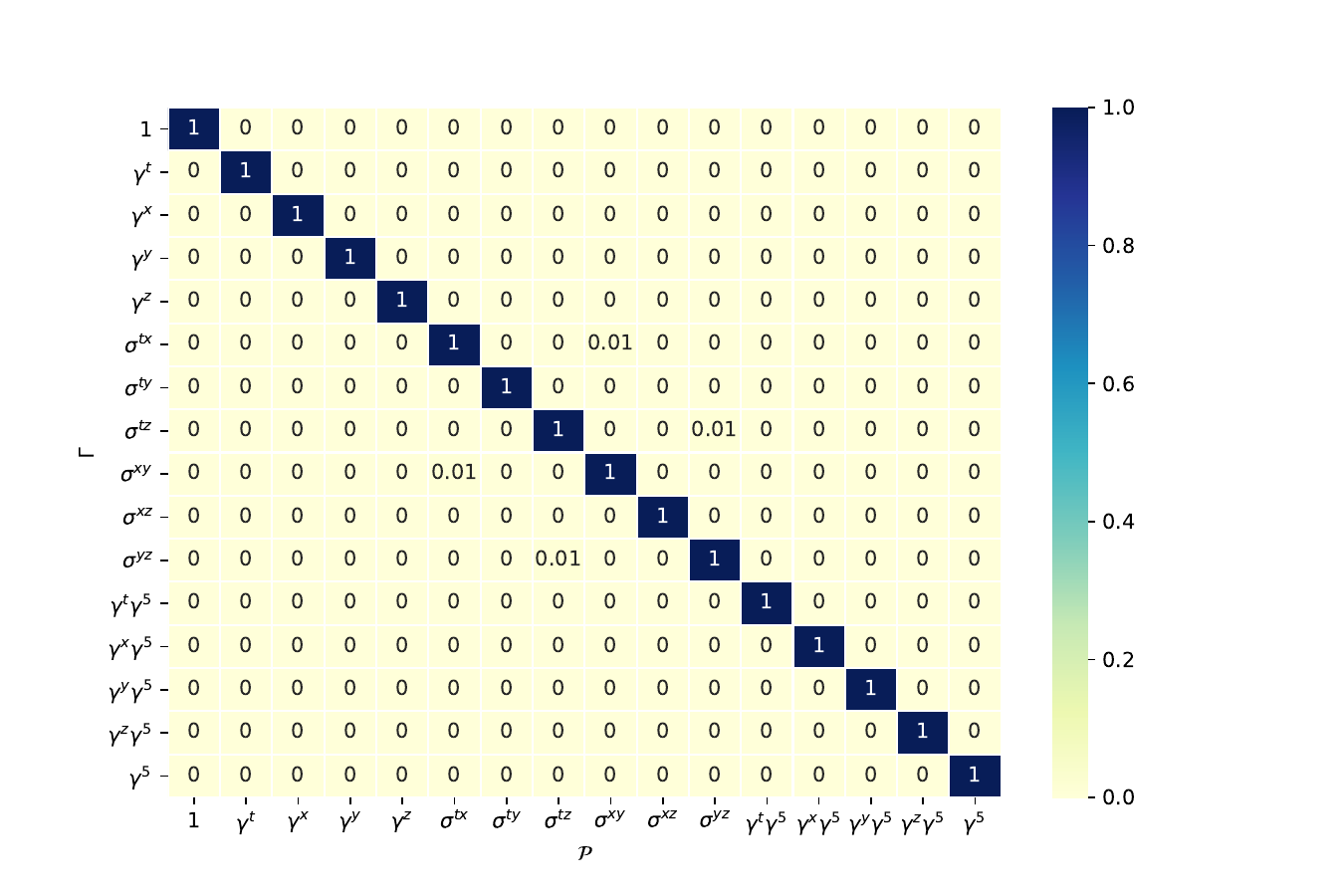}
\includegraphics[width=0.45\textwidth]{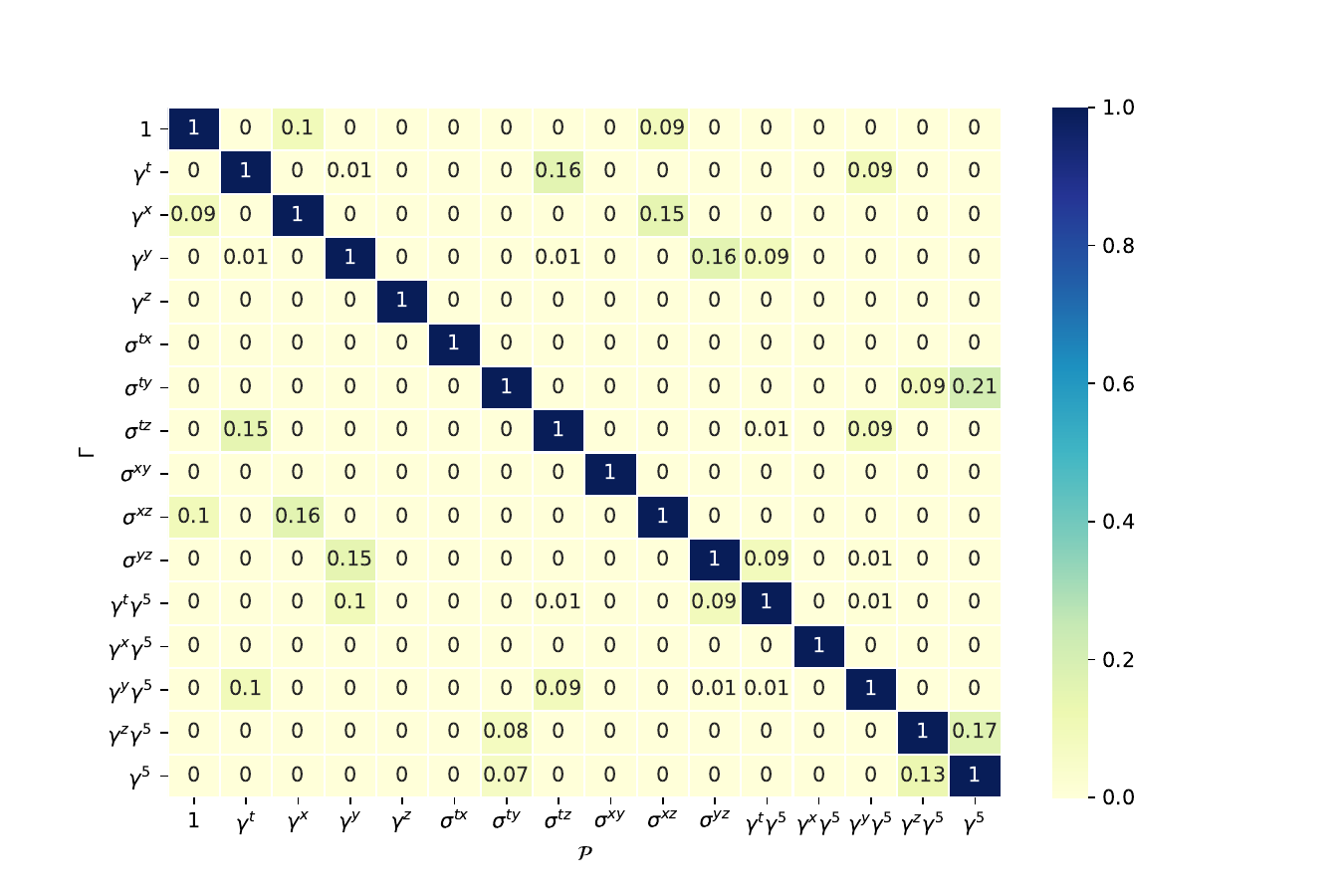}
\includegraphics[width=0.45\textwidth]{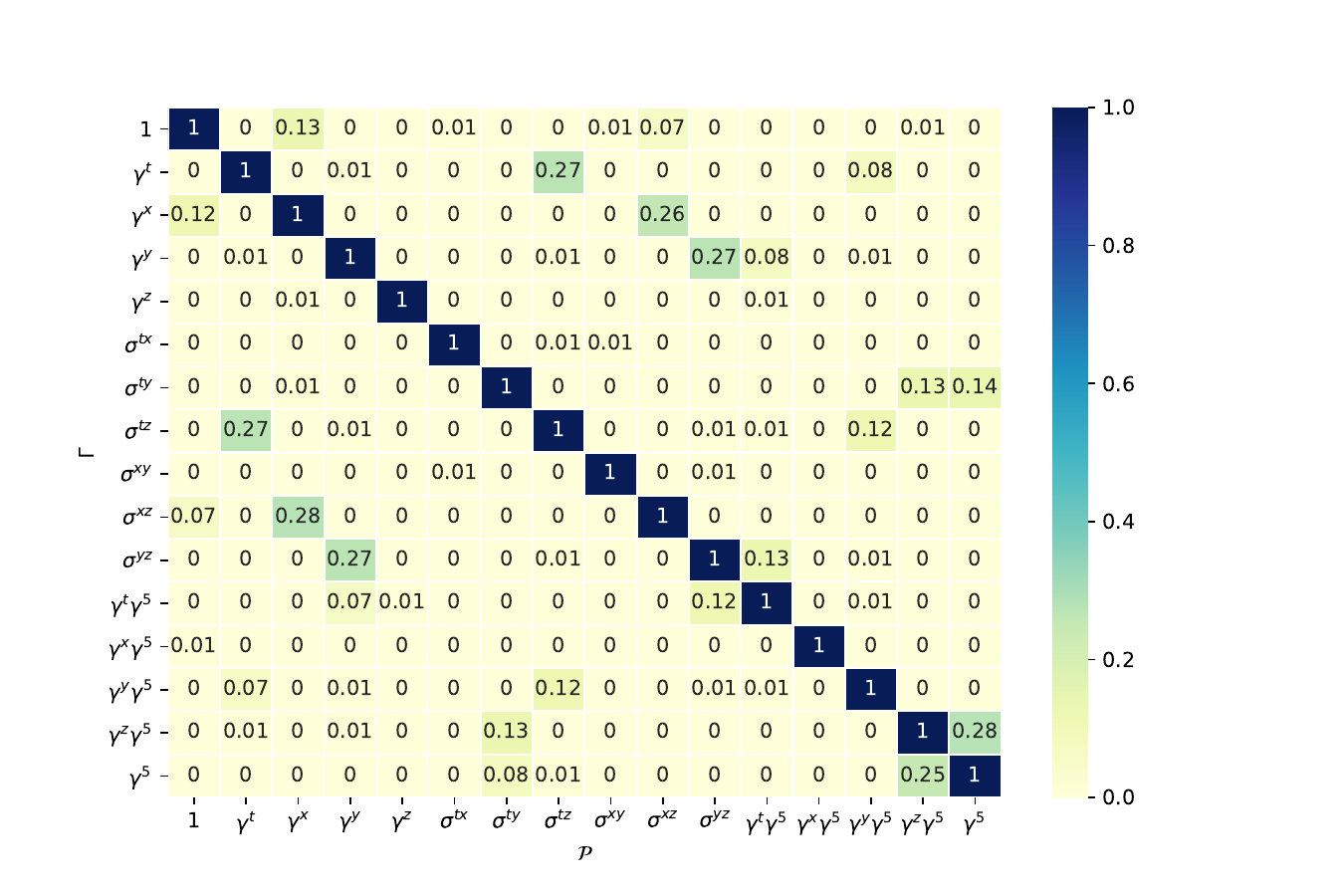}
\includegraphics[width=0.45\textwidth]{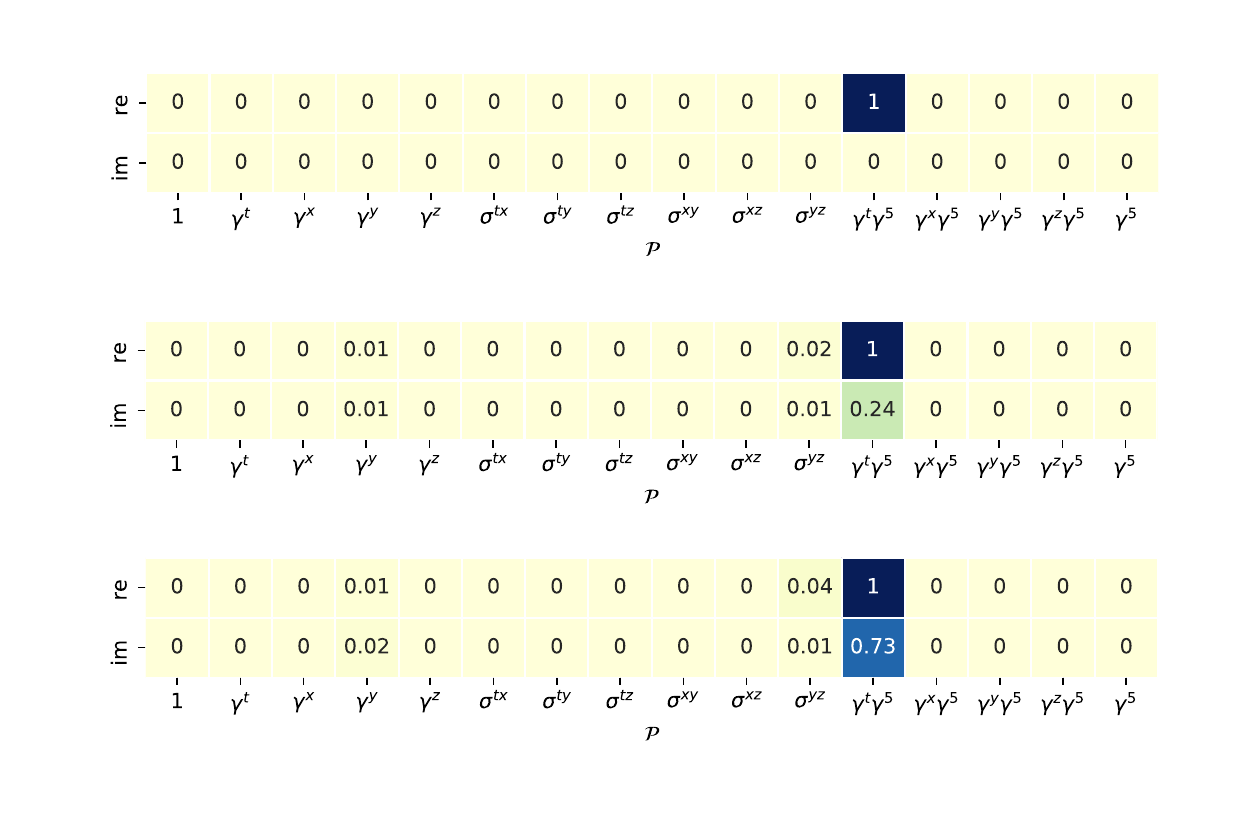}
\caption{{Heatmap for the nonperturbative renormalization factor with $P=1.58$GeV ($b_{\perp}=0$: upper left, $b_{\perp}=3a$: upper right, $b_{\perp}=6a$: lower left) and matrix elements (lower right, exhibited by the $b_{\perp}=\{0,~3a,~6a\}$ sequence). In this figure, $\Gamma$ indicates the Lorentz structure in the operators while ${\cal P}$ denotes the projection. }}\label{fig:HeatMap_NPR}
\end{center}
\end{figure}

\subsection{Estimate of Operator Mixing for TMDWF}

{
In order to estimate the operator-mixing effects,  we adopt the same method as Refs.~\cite{Shanahan:2019zcq,Shanahan:2020zxr} and calculate the nonperturbative RI/MOM renormalization/mixing factors,
\begin{align}
{\cal M}_{{\cal P}\Gamma}=\textrm{Tr}\left[{\cal P}\Big\langle q(p)\Big|\overline q_1\left(\frac{z }{2}n^z +\vec b\right)\Gamma \,{\cal W}(\vec b,\ell)q_2\left(-\frac{z}{2}n^z \right)\Big| q(p)\Big\rangle\right],
\end{align}
where $\Gamma$ indicates the Lorentz structure in the operators while ${\cal P}$ denotes the projection. The relative mixing effect is considered using the ratio ${\cal M}_{{\cal P}\Gamma}/{\cal M}_{\Gamma\Gamma}$. The results with three transverse separations $b_\perp=(0a, 3a, 6a)\hat{n}_x$ and off-shell quark momentum $p=(p_t,p_x,p_y,p_z)=(3.16,0,1.58,0)$ GeV are shown with the heatmap in Fig.~\ref{fig:HeatMap_NPR} (upper left, upper right and lower left panels correspondingly). The mixing effect grows with increasing $b_\perp$. This pattern is   consistent with the perturbative calculation in Ref.~\cite{Shanahan:2019zcq}.}

{
The relative mixing effect in the quasi-TMDWF can be estimated through the product of the bare quasi-TMDWF with given Lorentz structure ${\cal P}$ and the corresponding mixing factor ${\cal M}_{\Gamma{\cal P}}$,
\begin{align}
\frac{\delta_{{\cal P}}\phi(z,b_\perp,P^z)}{\phi(z,b_\perp,P^z)}\simeq\frac{{\cal M}_{\Gamma_{\phi}{\cal P}}\Big\langle 0\Big|\overline q_1\left(\frac{z }{2}n^z +\vec b\right){\cal P}\,{\cal W}(\vec b,\ell)q_2\left(-\frac{z}{2}n^z \right)\Big| {\pi(\vec{P})}\Big\rangle}{\textrm{Re}[{\cal M}_{\Gamma_{\phi}\Gamma_{\phi}}\Big\langle 0\Big|\overline q_1\left(\frac{z }{2}n^z +\vec b\right){\Gamma_{\phi}}\,{\cal W}(\vec b,\ell)q_2\left(-\frac{z}{2}n^z \right)\Big| {\pi(\vec{P})}\Big\rangle]},
\end{align}
with $\Gamma_{\phi}=\gamma_5\gamma_t$.
We give the results in the lower right panel of Fig.~\ref{fig:HeatMap_NPR} for $b_\perp=(0a, 3a, 6a)\hat{n}_x$. From the figure, one can find that the operator-mixing effects can reach   order 5\% for the transverse separation $b\sim$ 0.6 fm, while it is less significant for smaller transverse separations. 
}



\subsection{Tabulated results for the intrinsic soft function}

\begin{figure}[!th]
\includegraphics[width=0.45\textwidth]{04_sf_perturb_sqrt2.pdf}
\includegraphics[width=0.45\textwidth]{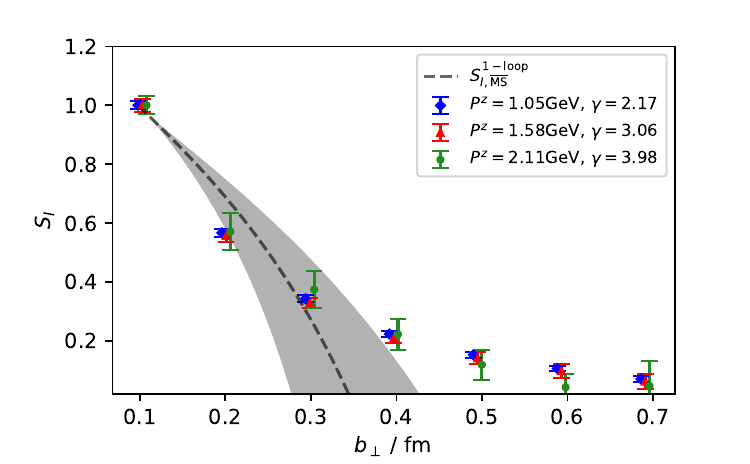}
\caption{ The intrinsic soft factor as a function of $b_{\perp}$  with  $b_{\perp,0}=a$ as {$S_{I,\overline{\rm MS}}(b_\perp,\mu)=\frac{F(b_\perp,P^z)}{F(b_{\perp,0},P^z)}\frac{|\phi(0,b_{\perp,0},P^z)|^2}{|\phi(0,b_\perp,P^z)|^2}+{\cal O}(\alpha_s, \gamma^{-2})$}. With different pion momentum $P^z$, the results are consistent with each other. The dashed  curve shows  the result of the 1-loop calculation, {$S_{I,\overline{\rm MS}}(b_{{\perp}},\mu)=1-\frac{\alpha_sC_F}{\pi}\ln\frac{\mu^2 b_{{\perp}}^2}{4 e^{-2\gamma_E}}+{\cal O}(\alpha_s)$},  with the strong coupling constant $\alpha_s(1/b_{\perp})$. {The shaded band corresponds to the scale uncertainty  of $\alpha_s$: $\mu\in [1/\sqrt{2},\sqrt{2}]\times 1/b_\perp$. The systematic uncertainty from the operator mixing has been taken into account. {Both the panels show the same data but the left panel uses the log scale and the right panel uses the normal one.}}}\label{fig:quasi-soft-factor2}
\end{figure}
 
{For ease of comparison as given in Fig.~\ref{fig:quasi-soft-factor2}, we give a tabulated results for the intrinsic soft function in Tab.~\ref{table:soft-function}.  The perturbative results for $b\le3a$ are consistent with our calculation taking into account the errors from the scale  {dependence} in the strong coupling constant $\alpha_s(\mu)$:  $\mu\in [1/\sqrt{2},\sqrt{2}]\times 1/b_\perp$.  }

\begin{table}[http]
\begin{center} 
\caption{Tabulated results for the intrinsic soft function as shown in Fig.~\ref{fig:quasi-soft-factor}. }\label{table:soft-function}
\begin{tabular}{|c|ccccccc|cc}
\hline
 & $b=a$         & $b=2a$         & $b=3a$         & $b=4a$         & $b=5a$         & $b=6a$         & $b=7a$         \\
\hline
$P=1.05$ GeV                   & 1.000(8)  & 0.567(7)  & 0.343(6)  & 0.224(6)  & 0.153(6)  & 0.106(7)  & 0.071(7)  \\
$P=1.58$ GeV     & 1.000(20) & 0.557(17) & 0.329(13) & 0.209(14) & 0.142(18) & 0.099(22) & 0.063(25) \\
$P=2.11$ GeV & 1.000(29) & 0.571(62) & 0.374(63) & 0.223(52) & 0.119(50) & 0.043(46) & 0.047(84)  \\
\hline
pQCD & $1.000_{-0.008}^{+0.005}$   &  $0.703_{-0.100}^{+0.057}$  &  $0.303_{-0.461}^{+0.193}$  & $-0.334_{-2.142}^{+0.498}$  & - & - & - \\
\hline
\end{tabular}
\end{center}
\end{table} 

\subsection{Two-state fit of the form factors}

In this work, we perform the following joint fit to obtain the norm of the subtracted quasi-TMDWF $|\phi_\ell(0,b_\perp,P^z,\ell)|$ and soft factor $S_I(b_\perp)$ (with $\ell=7a$),
\begin{align}
&\frac{C_3(b_\perp,P^z,t_{\rm sep},t)}{C_2(0,P^z,0,t_{\rm sep})}=\frac{|\tilde{\phi}_\ell(0,b_\perp,P^z,\ell)|^2\tilde{S}_I(b_\perp)+C_1(e^{-\Delta E t}+e^{-\Delta E (t_{\rm sep}-t)})+C_2e^{-\Delta E t_{\rm sep}}}{1+C_0 e^{-\Delta E t_{\rm sep}}}, \nonumber\\
&\frac{C_2(b_\perp,P^z,0,t)}{C_2(0,P^z,0,t)}=\frac{|\tilde{\phi}_\ell(0,b_\perp,P^z,\ell)|e^{\theta(b_\perp,P^z,\ell)}(1+C_3e^{-\Delta E t})}{1+C_0e^{-\Delta E t}},
\label{eq:fit}
\end{align}
where
\begin{align}
\tilde{\phi}_\ell(0,b_\perp,P^z,\ell)=\frac{\phi_\ell(0,b_\perp,P^z,\ell)}{\phi_\ell(0,0,P^z,0)},\ \tilde{S}_I(b_\perp)=\frac{{\phi}_L(0,0,P^z,0)A_w}{2EA_p}S_I(b_\perp),
\end{align}
and $\theta(b_\perp,P^z,\ell)$ is the phase of the  quasi-TMDWF.
The additional factor in the definition of $\tilde{S}_I$ will be cancelled by $\tilde{S}_I(b_\perp=a)$ {when we consider the following ratio,
\begin{align}\label{eq:S_ratio}
S_{I,\overline{\rm MS}}(b_\perp,\mu)= \left(\frac{S_I(b_{\perp},1/a)}{S_I(b_{\perp,0},1/a)}\right)S_{I,\overline{\rm MS}}(b_{\perp,0},\mu).
\end{align}}

\begin{figure}[!th]
\begin{center}
\includegraphics[width=0.45\textwidth]{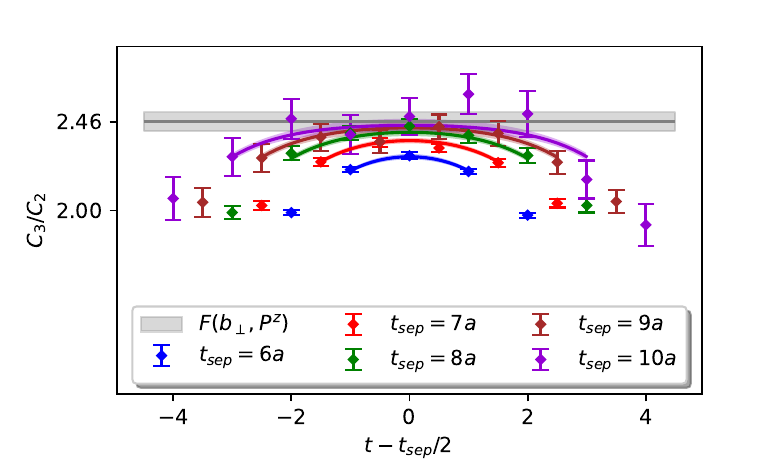}
\includegraphics[width=0.45\textwidth]{03_ff_P3b3.pdf}\\
\includegraphics[width=0.45\textwidth]{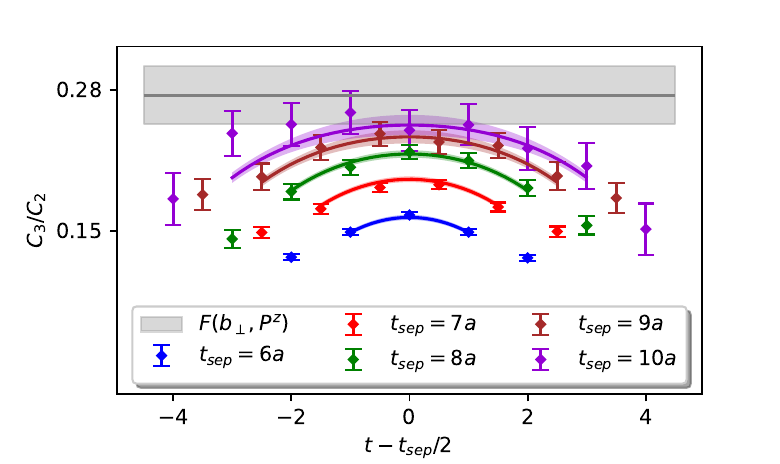}
\includegraphics[width=0.45\textwidth]{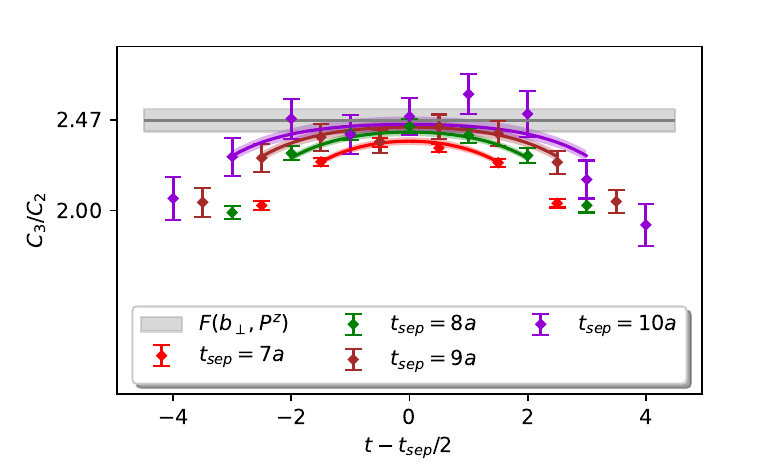}
\caption{  The ratios $C_3(b_\perp,P^z,t_{\rm sep},t)/C_2(0, P^z,0,t_{\rm sep})$ as   function of $t_{\rm sep}$ and $t$, with $\{P^z, b_{\perp}\}=\{6\pi/L, 1a\}$ (upper left panel), $\{P^z, b_{\perp}\}=\{6\pi/L, 3a\}$  (upper right panel) and $\{P^z, b_{\perp}\}=\{6\pi/L, 5a\}$ (lower left panel). {In the lower right panel, we have dropped the $t_{\rm sep}=6$a data for the case  with $\{P^z, b_{\perp}\}=\{6\pi/L, 1a\}$ and find that the fitted result  {is consistent with the case in the upper left panel.} }}\label{fig:joint_fit2}
\end{center}
\end{figure}

In Fig.~\ref{fig:joint_fit2}, we {show} the ratios $C_3(b_\perp,P^z,t_{\rm sep},t)/C_2(0, P^z,0,t_{\rm sep})$ with $P^z=6\pi/L$, $b_{\perp}=\{1a,3a,5a\}$, compared  with the two-state fit predictions (colored bands) and fitted ground state contribution (gray band). All of them show good agreement between data and fits.
{This agreement indicates that the systematic uncertainty from the fit-ranges is mild. As another estimate, we have dropped the $t_{\rm sep}=6$a data for the case  with $\{P^z, b_{\perp}\}=\{6\pi/L, 1a\}$ and the results are shown in  the lower right panel of  Fig.~\ref{fig:joint_fit2}. One can  find that the fitted result  {is   consistent  with the   case in the upper left panel within uncertainties.} }

\begin{figure}[!th]
\begin{center}
\includegraphics[width=0.7\textwidth]{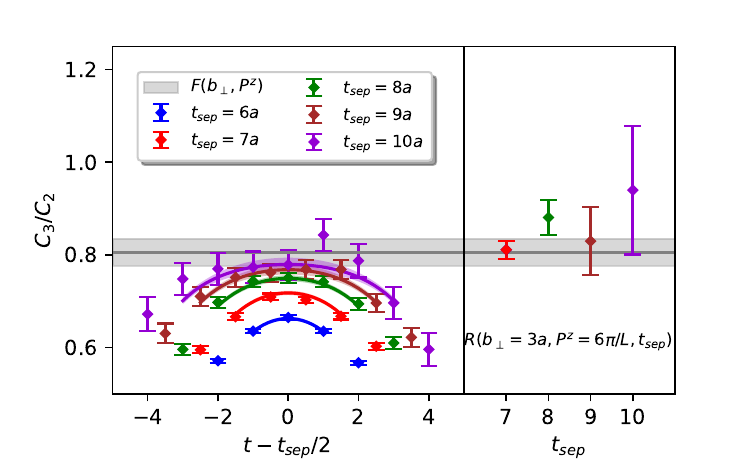}
\caption{ The ratios $C_3(b_\perp,P^z,t_{\rm sep},t)/C_2(0, P^z,0,t_{\rm sep})$ as   function of $t_{\rm sep}$ and $t$ compared with the differential summed  ratio $R(b_\perp,P^z,t_{\rm sep})$, with $\{P^z, b_{\perp}\}=\{6\pi/L, 3a\}$.}\label{fig:joint_fit3}
\end{center}
\end{figure}

As another check, we also consider the differential summed  ratio
\begin{align}
R(b_\perp,P^z,t_{\rm sep})&\equiv SR(b_\perp,P^z,t_{\rm sep})-SR(b_\perp,P^z,t_{\rm sep}-1)=|\tilde{\phi}_\ell(0,b_\perp,P^z,\ell)|^2\tilde{S}_I(b_\perp)+{\cal O}(e^{-\Delta E t_{\rm sep}}),\ \nonumber\\
SR(b_\perp,P^z,t_{\rm sep})&\equiv \sum_{0<t<t_{\rm sep}} \frac{C_3(b_\perp,P^z,t_{\rm sep},t)}{C_2(0, P^z,0,t_{\rm sep})}.
\end{align}
As an example, we plot $R(b_\perp,P^z,t_{\rm sep})$ as function of $t_{\rm sep}$ in Fig.~\ref{fig:joint_fit3} for $\{P^z, b_{\perp}\}=\{6\pi/L, 3a\}$ and compare it with the standard two-state fit. We can see that the $R(b_\perp,P^z,t_{\rm sep})$ agree with the ground state contribution from the two state fit at large $t_{\rm sep}$.

\subsection{The possible imaginary part in extracting the Collins-Soper kernel}

\begin{figure}[!th]
\begin{center}
\includegraphics[width=0.44\textwidth]{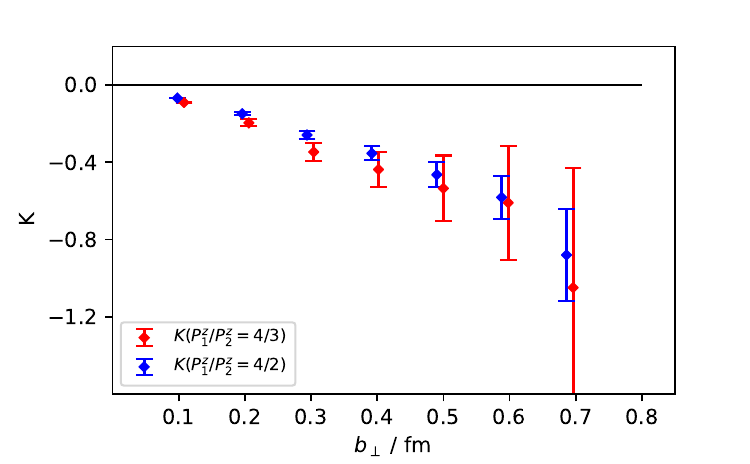}
\includegraphics[width=0.44\textwidth]{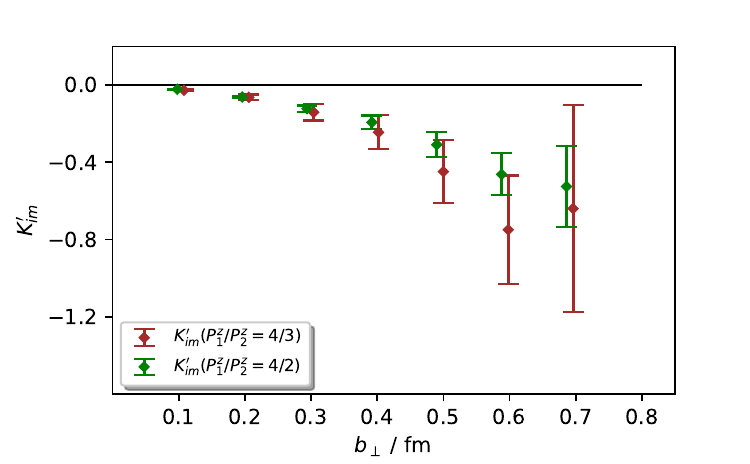}
\caption{ The real (left panel) and imaginary (right panel) parts of the Collins-Soper kernel when the approximation $C(xP^z,\mu)=1+{\cal O}(\alpha_s)$ is taken, based on the definition in Eq.~(\ref{eq:CSK_full}).}\label{fig:imaginal}
\end{center}
\end{figure}

\begin{figure}[!th]
\begin{center}
\includegraphics[width=0.45\textwidth]{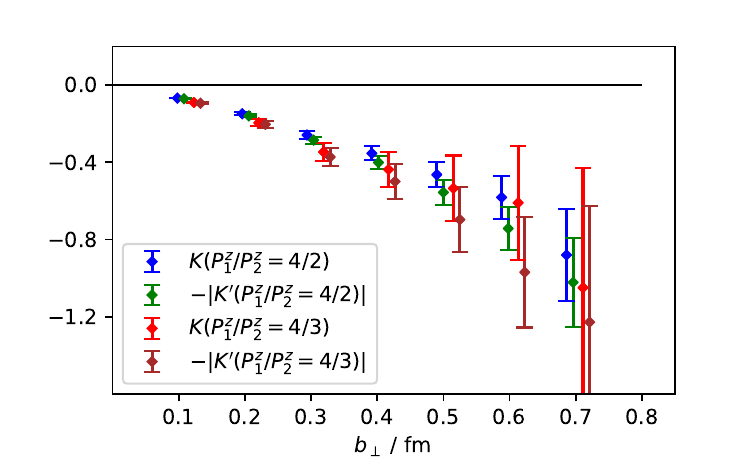}
\caption{The comparison on $K$ and $-|K'|=-\sqrt{K^2+K'^2_{\rm Im}}$. They are consistent with each other.}\label{fig:compare}
\end{center}
\end{figure}

The Collins-Soper kernel with the following definition
\begin{align}
&K'(b_\perp,\mu)=\frac{1}{\ln(P_1^z/P_2^z)}\ln\frac{C(xP_2^z,\mu) \Phi_{\overline{\rm MS}}(x,b_\perp,P_1^z,\mu)}{C(xP_1^z,\mu) \Phi_{\overline{\rm MS}}(x,b_\perp,P_2^z,\mu)}+{\cal O}(\gamma^{-2})=\frac{1}{\ln(P_1^z/P_2^z)}\ln\frac{\phi(0,b_\perp,P_1^z)}{\phi(0,b_\perp,P_2^z)}+{\cal O}(\alpha_s, \gamma^{-2})\, \nonumber\\
&\quad\quad\quad\ \ =\frac{1}{\ln(P_1^z/P_2^z)}\ln\left|\frac{\phi(0,b_\perp,P_1^z)}{\phi(0,b_\perp,P_2^z)}\right|+\frac{i}{\ln(P_1^z/P_2^z)}\left(\theta(0,b_\perp,P_1^z)-\theta(0,b_\perp,P_2^z)\right)+{\cal O}(\alpha_s, \gamma^{-2})\, \label{eq:CSK_full}
\end{align}
should be real, but the $P^z$ dependence of the phase $\theta(0,b_\perp,P^z)=\textrm{tan}^{-1}\frac{\phi_{\rm Im}(0,b_\perp,P^z)}{\phi_{\rm Re}(0,b_\perp,P^z)}$ can introduce an {imaginary} part of $K'$ when the approximation $C(xP^z,\mu)=1+{\cal O}(\alpha_s)$ is employed. Fig.~\ref{fig:imaginal} shows the real and imaginary parts   as   functions of $b_\perp$ with two combinations of $P_1/P_2$. The real part $K'_{\rm Re}=K$ (left panel)  corresponds to the definition used in  the main text.   The non-vanishing  imaginary part (right panel)  reflects the systematic    uncertainty due to imprecise matching.  $K$ and $-|K'|=-\sqrt{K^2+K'^2_{\rm Im}}$ are  still consistent within the statistical uncertainty of $K$ as in Fig.~\ref{fig:compare}.

To estimate the effect of inaccurate matching, we consider $|K'_{\rm Im}|$ as a systematic uncertainty and {add} it with the statistical uncertainty of $K$ in quadrature.

\end{widetext}




\end{document}